\documentclass[superscriptaddress,preprint]{revtex4-1}

\usepackage{graphicx}% Include figure files
\usepackage{dcolumn}% Align table columns on decimal point
\usepackage{bm}% bold math
\usepackage{multirow}
\usepackage{color}
\usepackage{float}
\usepackage{lineno}

\begin{document}
 
\title{Reprogrammable magnonic logic in a multiferroic heterostructure via magnetoelectric coupling}

\author{Ping Che}\email{ping.che@cnrs-thales.fr}
\affiliation{Laboratoire Albert Fert, CNRS, Thales, Université Paris-Saclay, Palaiseau, France}
\author{Amr Abdelsamie}
\affiliation{Laboratoire Albert Fert, CNRS, Thales, Université Paris-Saclay, Palaiseau, France}
\author{Ádám Papp}
\affiliation{Faculty of Information Technology and Bionics, Pázmány Peter Catholic University, Budapest, Hungary}
\author{Sali Salama}
\affiliation{Centre de Nanosciences et de Nanotechnologies, CNRS, Université Paris-Saclay, Palaiseau, France}
\author{André Thiaville}
\affiliation{Laboratoire de Physique des Solides, CNRS, Université Paris-Saclay, Orsay, France}
\author{Romain Lebrun}
\affiliation{Laboratoire Albert Fert, CNRS, Thales, Université Paris-Saclay, Palaiseau, France}
\author{Stéphane Fusil}
\affiliation{Laboratoire Albert Fert, CNRS, Thales, Université Paris-Saclay, Palaiseau, France}
\author{Vincent Garcia}
\affiliation{Laboratoire Albert Fert, CNRS, Thales, Université Paris-Saclay, Palaiseau, France}
\author{Aymeric Vecchiola}
\affiliation{Laboratoire Albert Fert, CNRS, Thales, Université Paris-Saclay, Palaiseau, France}
\author{Karim Bouzehouane}
\affiliation{Laboratoire Albert Fert, CNRS, Thales, Université Paris-Saclay, Palaiseau, France}
\author{Manuel Bibes}
\affiliation{Laboratoire Albert Fert, CNRS, Thales, Université Paris-Saclay, Palaiseau, France}
\author{Agnès Barthélémy}
\affiliation{Laboratoire Albert Fert, CNRS, Thales, Université Paris-Saclay, Palaiseau, France}
\author{Jean-Paul Adam}
\affiliation{Centre de Nanosciences et de Nanotechnologies, CNRS, Université Paris-Saclay, Palaiseau, France}
\author{Vladislav Demidov}
\affiliation{Institute of Applied Physics, University of Münster, Münster, Germany}
\author{Paolo Bortolotti}
\affiliation{Laboratoire Albert Fert, CNRS, Thales, Université Paris-Saclay, Palaiseau, France}
\author{Abdelmadjid Anane} \email{madjid.anane@university-paris-saclay.fr}
\affiliation{Laboratoire Albert Fert, CNRS, Thales, Université Paris-Saclay, Palaiseau, France}
\author{Isabella Boventer}\email{isabella.boventer@cnrs-thales.fr}
\affiliation{Laboratoire Albert Fert, CNRS, Thales, Université Paris-Saclay, Palaiseau, France}

\date{\today}

\begin{abstract}
%No more than 150 words
The realization of fully reconfigurable, voltage-controlled, and programmable on-chip magnonic devices is essential to fully harness the potential of spin waves for signal processing, logic and neuromorphic computing. Yet, existing demonstrations of electrical tuning of magnonic responses are either volatile, current-driven and thus energy-inefficient, or rely on local strain modification limiting their scalability for wafer-scale integration. Here, we address this challenge using a BiFeO$_{3}$/La$_{0.67}$Sr$_{0.33}$MnO$_{3}$ multiferroic thin film heterostructure. We show that ferroelectric domain engineering in BiFeO$_{3}$ enables deterministic tuning of the magnon dispersion of La$_{0.67}$Sr$_{0.33}$MnO$_{3}$, producing frequency shifts up to $\sim$150 MHz and allowing reconfigurable waveguiding. Micro-focused Brillouin light scattering directly images these effects, revealing electrically defined magnonic waveguides and spatially programmable dispersion. Compared to conventional approaches, this method provides non-volatile and reversible control. Furthermore, using an inverse-design simulation code, we demonstrate the capability of our platform to perform advanced magnonic functions such as frequency demultiplexing. Our results open a new avenue for using magnetoelectric heterostructures for magnonic logic, with further applicability to reservoir and neuromorphic computing and AI driven magnonic devices.
\end{abstract}

\maketitle
%\linenumbers

\newpage

\section*{Introduction}
Reconfigurable magnonic circuits are expected to enable adaptive control of signal flow and logic functions in compact, low-power architectures~\cite{Grundler2015,Chumak2022,Flebus2025,Haldar2016,Wang2018,Merbouche2021,Chai2024,Zenbaa2025}. In these systems, information is carried by the collective excitations of magnetically ordered systems, spin waves and their quanta magnons, which are a manifestation of pure spin currents allowing information transport without net charge motion~\cite{Gurevich1996,Kruglyak2010,Chumak2015,Csaba2017,Bensmann2025}. This property enables intrinsically low-dissipation operation. Furthermore, spin waves couple efficiently to other quasiparticles such as phonons to light, i.e. photons, and the physical modes underlying qubit states, thereby expanding their potential for hybrid information manipulation, 
storage, and processing, from room temperature down to the millikelvin range~\cite{An2020,Xu2020,Tabuchi2014,Diederich2023,Tabuchi2015,Lachance-Quirion2020}. In this context, on-chip reconfigurability is emerging as a key requirement for magnonics to bridge the gap from laboratory prototypes to practical applications. It is therefore expected to play a central role in future magnonic technologies, where complex functions can be redefined on demand, within short timescales, and with minimal energy consumption.

A wide range of magnonic building blocks has already been demonstrated, including diodes, couplers, (de)multiplexers, conduits, majority gates, spin-wave amplifiers, and transistors~\cite{Wang2018,Lan2015,Wang2020,Heussner2020,Heinz2020,Fischer2017,Talmelli2020,Chumak2014,Wang2024,Cheng2024}. However, for scalable on-chip implementation of reprogrammable functions, devices must combine submicron dimensions, robust spin-wave propagation, and zero-standby-energy reconfigurability. One promising route to meeting these requirements is voltage control of the effective magnetic landscape experienced by propagating spin waves~\cite{Merbouche2021,Chai2024,Wang2024,Rovillain2010,Choudhury2020,Liu2021,Qin2021,Rana2024}. This can be achieved via a ferroelectric top layer, which enables non-volatile control through its polarization state~\cite{Cheng2024,Vermeulen2019}. A particularly attractive implementation, explored in the present work, employs a multiferroic material in which the ferroelectric and antiferromagnetic orders are intrinsically coupled, specifically BiFeO$_3$ (BFO)~\cite{Wang2003,Gross2017,Husain2025}. This widely studied multiferroic perovskite is notable for its strong ferroelectric polarization ($\sim$ 100 $\mu$C/cm$^2$), high ferroelectric ordering temperature ($T_{\text{C}}$ $\sim$ 1100 K), and cycloidal antiferromagnetic ordering temperature ($T_{\text{N}}$ $\sim$ 640 K). When BFO is coupled to a low-loss magnonic medium such as half-metallic La$_{0.67}$Sr$_{0.33}$MnO$_{3}$ (LSMO)~\cite{Liu2019,Zhang2025}, the resulting bilayer forms an ideal platform for electrically reconfigurable magnonic devices, combining low-damping propagation with direct voltage control via magnetoelectric coupling (ME)~\cite{Merbouche2021,Wu2010,Yu2010,Zhang2021}.

Here, using micro-focused Brillouin light scattering (BLS) spectroscopy, we observe a shift of  in the spin-wave dispersion relation in the LSMO layer between opposite BFO polarization states, attributable to changes in the effective internal magnetic field. By directly writing the ferroelectric polarization using the trailing field method with a biased tip on piezoresponse force microscopy (PFM), we redefine the micrometer-scale magnonic circuit layout at will. To demonstrate magnonic channeling of spin waves, we employ phase-resolved Brillouin-Light Scattering Spectroscopy (BLS), providing direct access to the propagating spin-wave's phase front modification through wavelength measurements. Micromagnetic simulations not only reproduce the experimental observations but also enable the inverse design of more complex functionalities~\cite{Zenbaa2025,Zenbaa2024,Wang2021,Papp2021}, exemplified here by a frequency demultiplexer. Our approach establishes a robust pathway to fully voltage-programmable, universal magnonic circuitry at the sub-micron scale, offering versatile architectures for room-temperature logic, signal processing, and neuromorphic computing, while remaining compatible with operation at cryogenic temperatures for hybrid quantum technologies.

\section*{Results}

\subsection*{Multiferroic heterostructure devices}

\begin{figure}
	\centering
	\includegraphics[width=1 \columnwidth]{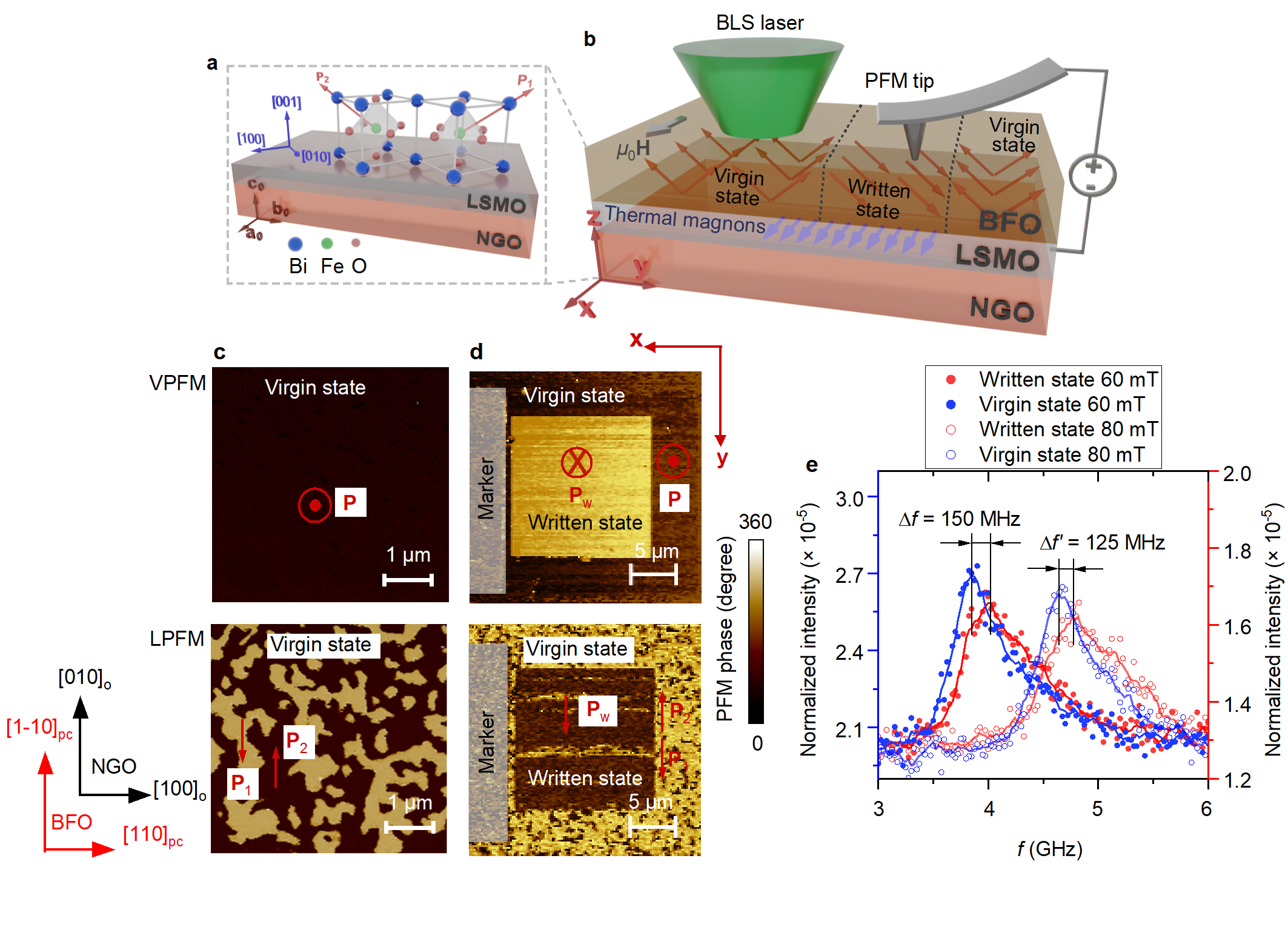}
	\caption{\textbf{System and thermal magnon spectra of La$_{0.67}$Sr$_{0.33}$MnO$_{3}$ (LSMO) in regions with virgin and written ferroelectrical states in BiFeO$_3$.} \textbf{a}, Schematic diagram of the epitaxially grown BFO on  LSMO and (001)$_\mathrm{o}$-NdGaO$_3$ (NGO) substrate. Green arrows indicate the crystalline orientation of BFO. $\textbf{a}_\mathrm{o}$, $\textbf{b}_\mathrm{o}$, and $\textbf{c}_\mathrm{o}$ represent the orthorhombic axes of the NGO. Red arrows $\textbf{P}_1$ and $\textbf{P}_2$ show the ground state of the ferroelectric (FE) polarization directions. \textbf{b}, Illustration of scanning Brillouin light scattering (BLS) detecting thermal magnons in LSMO and FE domains writing in BFO layer. The electric field was applied between the PFM tip and the LSMO layer serving as bottom electrode for FE domain writing. In the written region, the FE polarization is switched from $\textbf{P}_1$ and $\textbf{P}_2$ to $\textbf{P}_\mathrm{w}$ (opposite direction of $\textbf{P}_2$) as marked by the red arrows. \textbf{c},\textbf{d}, Vertical PFM (VPFM) and lateral PFM (LPFM) images of the FE domains in the virgin states and written states as in \textbf{b}. \textbf{e}, Thermal spectra of magnons in LSMO in the region corresponding to written and virgin states in \textbf{d}. All spectra are averaged scans over four locations normalized to the elastic peaks of each BLS scan. Solid curves are the adjacent-averaging of the BLS data.}
	\label{Thermal_magnon}
\end{figure}

The heterostructure of BFO/LSMO is epitaxially grown by pulsed laser deposition (PLD) on (001)$_\mathrm{o}$-oriented NGO substrates, as illustrated in Fig. 1$\textbf{a}$. The growth parameters are described in detail in the Methods section. All magnonic devices investigated in this work are comprised of a  BFO layer of 34 nm and LSMO layer of 21 nm thickness, respectively. In this configuration, BFO and LSMO adopt the (001)$_\mathrm{pc}$ orientation. The orthorhombic nature of (001)$_\mathrm{o}$-NGO  substrate is indicated by the axes $\textbf{a}_\mathrm{o}$, $\textbf{b}_\mathrm{o}$ and $\textbf{c}_\mathrm{o}$ in Fig. 1\textbf{a}. It imposes anisotropic strain in both BFO and pseudocubic LSMO layers, modifying the ferroic properties of BFO film compared to their bulk counterparts~\cite{Abdelsamie2024}. The in-plane magnetization of LSMO layer exhibits strong uniaxial anisotropy. Correspondingly, the ferroelectric domain configuration of the BFO film is reduced from eight variants in bulk to two variants separated by 109$^{\circ}$ (represented by red arrows $\textbf{P}_1$ and $\textbf{P}_2$ in Fig. 1\textbf{a}). We characterized both the out-of-plane and the in-plane ferroelectric components of the as-grown sample, labeled as virgin state, by vertical and lateral piezoresponse force microscopy (PFM) imaging. Fig. 1\textbf{c} shows that the out-of-plane ferroelectric components are uniformly up-polarized (dark brown color), whilst the in-plane components exhibits the two different orientations (light and dark brown color) corresponding to $\textbf{P}_1$ and $\textbf{P}_2$ in Fig. 1\textbf{a}.

In order to deterministically and reversibly control the orientation of the ferroelectric polarization vector in the BFO layer, we employed the trailing field method using a biased PFM tip \cite{Gross2017}. This approach leverages the synergetic tip scanning direction and bias to simultaneously manipulate both the out-of-plane and in-plane polarization components. Fig. 1\textbf{d} (top panel) exemplarily displays the vertical PFM image of well-defined, square-shaped switched ferroelectric domain of size 15 $\times$ 15 $\mu$m$^2$ (labeled as written state) surrounded by the as-grown FE domain configuration. Correspondingly, the lateral PFM in bottom panel of Fig. 1\textbf{d} clearly reveals the in-plane polarization switching, as indicated by the brown box. Using this approach, the ferroelectric polarization states are controllably reversed from an initial two-variant configuration ($\textbf{P}_1$ and $\textbf{P}_2$) into predominantly single-domain state ($\textbf{P}_{\text{w}}$), a configuration that is crucial for our devices. The top-panel of Fig.\ref{Thermal_magnon} (d) displays the out-of-plane and the bottom panel the in-plane component of \textbf{P}. The FE polarization $\textbf{P}_\mathrm{w}$ is indicated by red arrows in the written state in Fig. 1\textbf{b}.\\
\indent
In order to assess the static and dynamic magnetic properties of our LSMO layer, we performed ferromagnetic resonance  (FMR) measurements at room temperature to extract the effective magnetization $M_\mathrm{eff} = (336 \pm 6)$ mT in the virgin state and the Gilbert damping $\alpha=5.9 \cdot 10^{-3}$ using Kittel formula (see Supplementary Information Section S3)~\cite{Kittel}. In combination with the inhomogeneous broadening of $\approx 2.5\,\mathrm{mT}$, these values indicate a good magnetic homogeneity of the LSMO film and are comparable to reported values  \cite{Merbouche2021,Zhang2021}.

\subsection*{Reproducibility of the ferroelectric domain switching}
%\subsection*{reprogammable }

\begin{figure}
    \centering
    \includegraphics[width=0.5 \columnwidth]{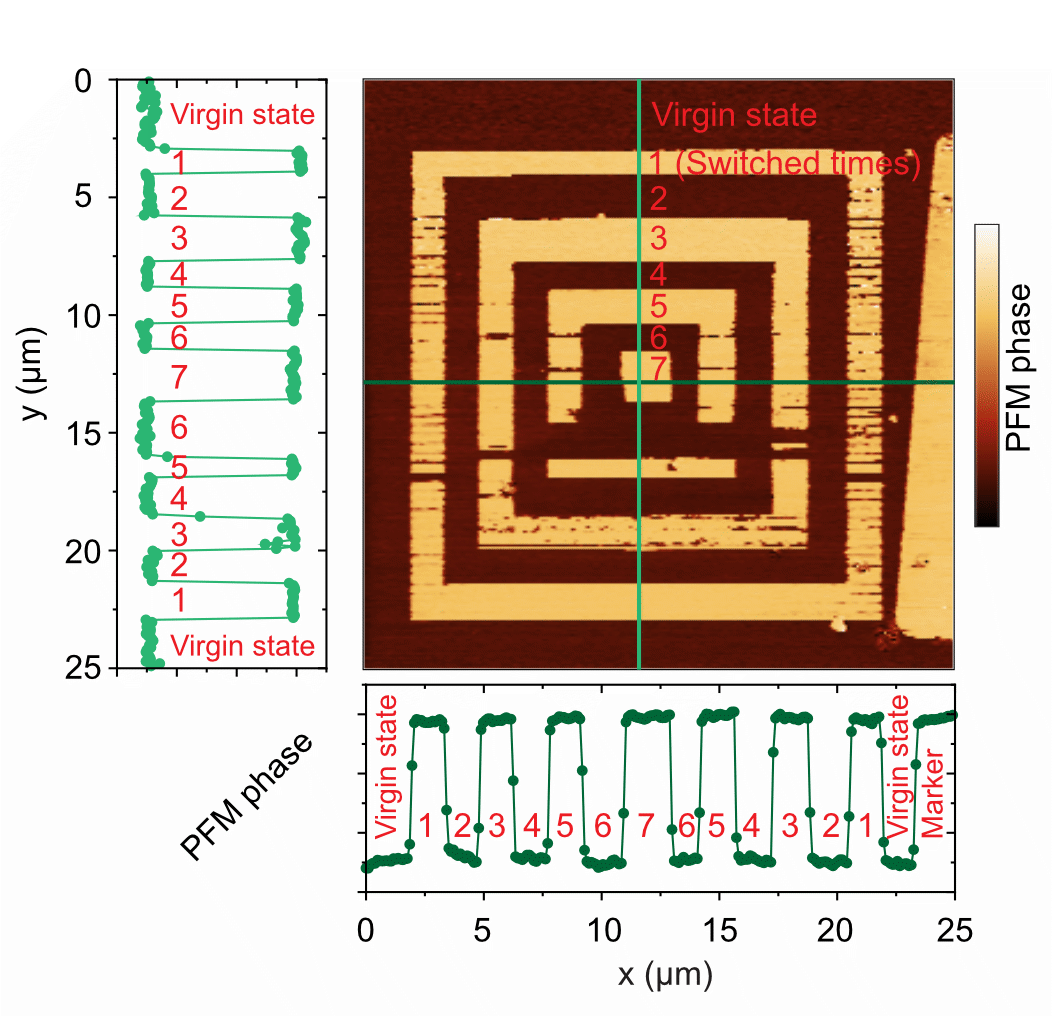}
    \caption{\textbf{Reproducibility of the ferroelectric domain writing in BFO layer.} 
    %\textbf{a},
    LPFM image of ferroelectric domains switched multiple times. The switching time of the domain are labeled by red-color numbers. Lineplots of the PFM voltage at the position of the light green and green lines are plotted on the left and bottom panel. 
    %\textbf{b}, Zoomed-in LPFM image of the region corresponding to the blue frame in \textbf{a} and the cycloid/antiferromagnetic domains in this region detected by the NV-center magnetometry.
    }
    \label{Reproducibility}
\end{figure}

Reprogrammability is a key property for both magnonic and multiferroic devices~\cite{Guo2013,Heron2014}. The cycling stability and endurance of reprogramming FE polarization for magnonic devices is critical and not yet well-tested~\cite{Rana2019}. By optimizing the growth process, we achieved high-quality control of the switching behavior using trailing field method. Fig. ~\ref{Reproducibility} shows a vertical PFM image of one representative region subjected to multiple switching cycles. Square regions of different size ranging from 2 to 20 $\mu$m were switched between one and seven times. In the dark green and light green linecuts, the PFM signal maintains the same phase even after seven switching cycles, indicating not only robust but also reproducible FE domain formation. Notably, all written and repeatedly switched FE domains remain stable over two years.Thus, they maintain the identical states throughout this timespan and demonstrate the reliability of the BFO/LSMO multiferroic heterostructure for device applications. Additionally, NV center-magnetometry performed in the virgin state of BFO, allowed to characterize the cycloid state in the BFO layer (FIG S2). This indicates a high quality of our thin layer.  

\subsection*{Internal field modification}
%from Ping: I think we should avoid to use the previous published work as a starting point here. this would weaken the novelty of our work.
%Previous works already reported on a modulation of the spin waves in LSMO by imprinting a magnonic crystal via the FE domain state (out-of plane only, in-plane not precisely defined) but the direct impact on the spin wave propagation has not been studied yet. Here, we are interested to further study key performance properties of a reprogrammable multiferroic device such as its ability to provide spin wave guiding within nm-sized channels of the same orientation of the ferroelectric polarization \textbf{P} in the BFO layer. Thus, after ensuring the reconfigurability of our system, we examined  
The dynamic magnetic properties of the LSMO layer corresponding to different FE domains via thermally excited magnons were examined. For this purpose, we employed micro-focus BLS ($\mu$BLS) with a 532 nm-wavelength green laser source for thermal magnon detection (see Methods section for details). Its scanning function and 300 nm laser spot size allow to probe the local magnonic properties in the virgin (blue points) and written states (red points), as shown in Fig.~\ref{Thermal_magnon} \textbf{e}. In order to verify the generic behaviour for different resonance frequencies (e.g. center FMR frequency corresponding to the signal peak ), we applied two different external fields of $\mu_0H$ = 60 mT (solid points in Fig. \ref{Thermal_magnon} (e)) and 80 mT (empty points in Fig. \ref{Thermal_magnon} (e)), respectively. The corresponding thermal spectra are displayed in Fig.~\ref{Thermal_magnon}\textbf{e}. Each spectrum is an average over four measurement points spaced 1 $\mu$m apart, confirming good spatial uniformity.\\
\indent
Whilst the peak positions indicate the FMR frequencies, the spectral tails extending toward higher (lower) frequencies correspond to magnetostatic surface (backward volume (BV)) magnetostatic magnon modes. Frequency shifts of $\Delta f$ = 150 MHz at 60 mT and $\Delta f'$ = 125 MHz at 80 mT are observed when comparing the peak positions. At 60 mT, this corresponds to an internal field difference of 3.5 mT.

In addition, this value corresponds to 1.33 of the frequency linewidth of the pure FMR signal at 5 GHz (Supplementary Materials Fig. S1) and there is no direct overlap between both signals if channeled at either frequeny.   Thus, this represents a significant frequency shift which is large enough to enable the design of magnonic logic devices by controlling the spin wave propagation via magnetoelectric coupling. 

%These frequency shifts are reflected across the entire spectra, indicating that the FE domains influence not only the FMR but also the dispersion relations of surface and volume magnon modes. The consistent shift across the full curves demonstrates that the effect is robust and significant enough to tune magnon behavior. For instance, it could be used to guide coherently excited spin waves along dedicated lines defined only by the specific orientation of $\textbf{P}$. 

The underlying mechanism is attributed to the possible combined effect of local exchange bias  between the ferromagnetic order of LSMO and the antiferromagnetic order of the multiferroic BFO originating from orbital reconstruction and charge transfer between Fe and Mn ~\cite{Merbouche2021,Yi2019}, charge screening i.e. charge injection/ depletion in LSMO induced by the FE polarization \cite{Huang2018,Gradauskeite2024} or oxygen migration at the interface ~\cite{Kim2014,Guo2018}. 
Based on our k-resolved BLS measurements, we can  exclude the presence of interfacial DMI (see also Fig. S4). However, to pin down the driving mechanism in this system requires a further independent materials science study on itself which is out of the scope of this work but we believe deserves definitely investigation in the future.

We will focus on the induced joint magnonic effect from all these interactions in this article.

\subsection*{Electrically tuned magnonic waveguides}
%Link with the previous section missing (from Agnes)

\begin{figure}
    \centering
    \includegraphics[width=\columnwidth]{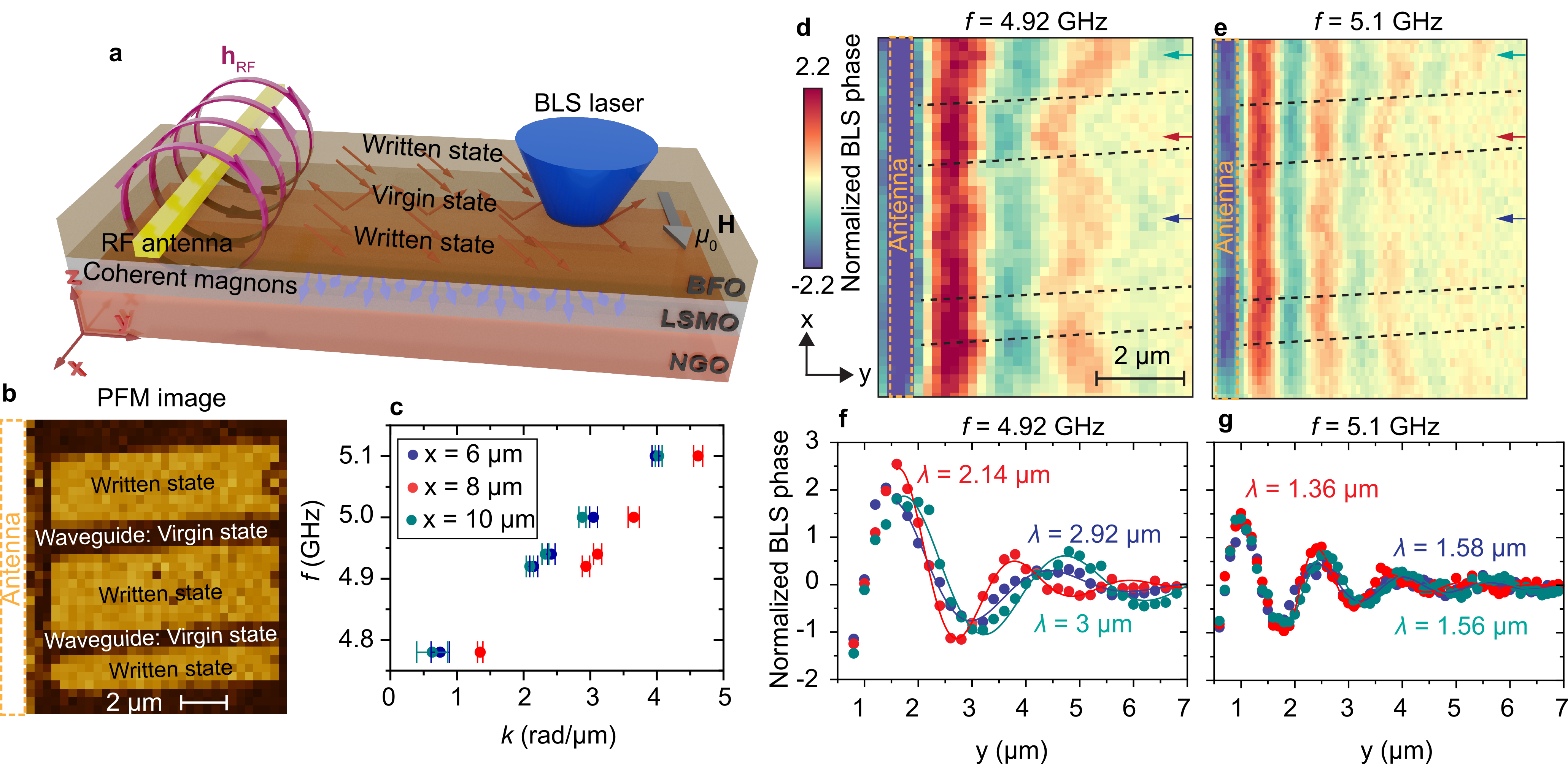}
    \caption{\textbf{Phase-resolved BLS spectra of RF-excited magnons in LSMO layer in the region of the imprinted waveguide.} \textbf{a}, Schematic diagram of the two-dimensional (2D) BLS scan over the region of magnonic waveguide in LSMO layer imprinted by the ferroelectric domains (red arrows) in BFO layer. A single-line antenna is used for RF injection and $\textbf{h}_{RF}$ induces the coherent magnons at GHz frequencies in LMSO. An external magnetic field is applied along the x-axis. \textbf{b}, PFM image of the ferroelectric domains in BFO layer and the magnonic waveguide region in LSMO. The region corresponding to virgin state is used as waveguide and the written states shape the edge of the waveguide. \textbf{c}, Magnon dispersion relations of LSMO layer extracted from the phase-resolved BLS spectra modified by ferroelectric domains in the BFO layer. \textbf{d}, \textbf{e}, Two-dimensional phase-resolved BLS spectra in the region shown in \textbf{b}. Dashed black lines mark the boundaries between the written state and the virgin state (waveguides). \textbf{f}, \textbf{g}, Circles are the one-dimensional scan along y-axis of phase-resolved BLS spectra at the location marked by the arrows in \textbf{d} and \textbf{e}. Solid curves are the fitted spin wave profiles using the damped Sine function, from which the wavelength is extracted. In \textbf{c}, \textbf{f} and \textbf{g}, red, blue and green circles are corresponding to the arrows in red (blue and green) in \textbf{d} and \textbf{e}, which are in the virgin (written) states.}
    \label{Propagate_magnon}
\end{figure}

Stripeline-shaped magnonic waveguides ( 1 $\mu$m width) were defined perpendicular to the prefabricated 200 nm-wide Au/Ti single-line antenna, as shown in the sketch in Fig.~\ref{Propagate_magnon}\textbf{a} and the PFM image in FIG. \textbf{b}. Since the switched FE domains in BFO exhibit higher resonance frequencies than the virgin FE domains, the virgin state is used as the waveguide to achieve better magnon confinement. The surrounding areas were switched to the written state using voltages applied to the PFM tip, as described in the previous sections. Red arrows in the BFO layer in Fig.~\ref{Thermal_magnon}\textbf{a} indicate the FE polarizations. PFM switching was performed very close to the antenna, with a gap of approximately 0.7 $\mu$m, avoiding any contact or damage to the antenna during the writing process. Magnetoelectric coupling in the BFO/LSMO heterostructure is an interfacial effect, so the study focused on magnetostatic surface magnon modes, whose propagating direction is perpendicular to the magnetization ($\textbf{k}\perp\textbf{M}$)~\cite{Gurevich1996}. An external field $\mu_0H$ = 80 mT was applied parallel to the antenna to saturate the LSMO layer and ensure the surface character of the magnon modes. Dynamic field $\textbf{h}_\mathrm{rf}$ from the injected radio-frequency current in the antenna excites coherent magnons in the LSMO layer in the range of 4.7 GHz to 5.1 GHz. 

We used the phase-resolved function of the $\mu$BLS with a blue laser source to obtain the dispersion relations of magnons inside and outside the waveguide. The extracted magnon dispersion relation is shown in Fig.~\ref{Propagate_magnon}\textbf{c}, where a downshift is observed in the waveguide region compared to the surrounding areas. To visualize this behavior in real space, we present two-dimensional (2D) magnon phase maps encoded with intensity at $f$ = 4.92 GHz and 5.1 GHz in Fig.~\ref{Propagate_magnon}\textbf{d} and \textbf{e}. The dashed lines indicate the boundaries between virgin and written states in the location corresponding to \textbf{b}. Phase fronts vary inside the waveguides, where magnons propagate with shorter wavelength $\lambda$. To quantitatively analyze the $\lambda$ variation, we extracted the phase profile along the y-direction at the positions marked by arrows in Fig.~\ref{Propagate_magnon}\textbf{d} and \textbf{e}. These profiles are shown in Fig.~\ref{Propagate_magnon}\textbf{f} and \textbf{g}, along with decayed sine function fits used to extract the $\lambda$. The red dots and fitted curves (inside the waveguide) correspond to shorter $\lambda$ than the green and blue ones (outside the waveguide), consistent with overall trend summarized in Fig.~\ref{Propagate_magnon}\textbf{c}.
Finally, we explore the potential to harness our multiferroic/ferromagnetic BFO/LSMO heterostructures for future magnonic devices. By using  the inverse design method we simulate a frequency demultiplexer (Fig. \ref{Simulation_Waveguides})~\cite{Heussner2020}. The parameters are chosen such that the structure could be immediately realized experimentally. 

\subsection*{Inverse-design magnonics devices via magnetoelectric coupling}

\begin{figure}
    \centering
    \includegraphics[width=0.95\columnwidth]{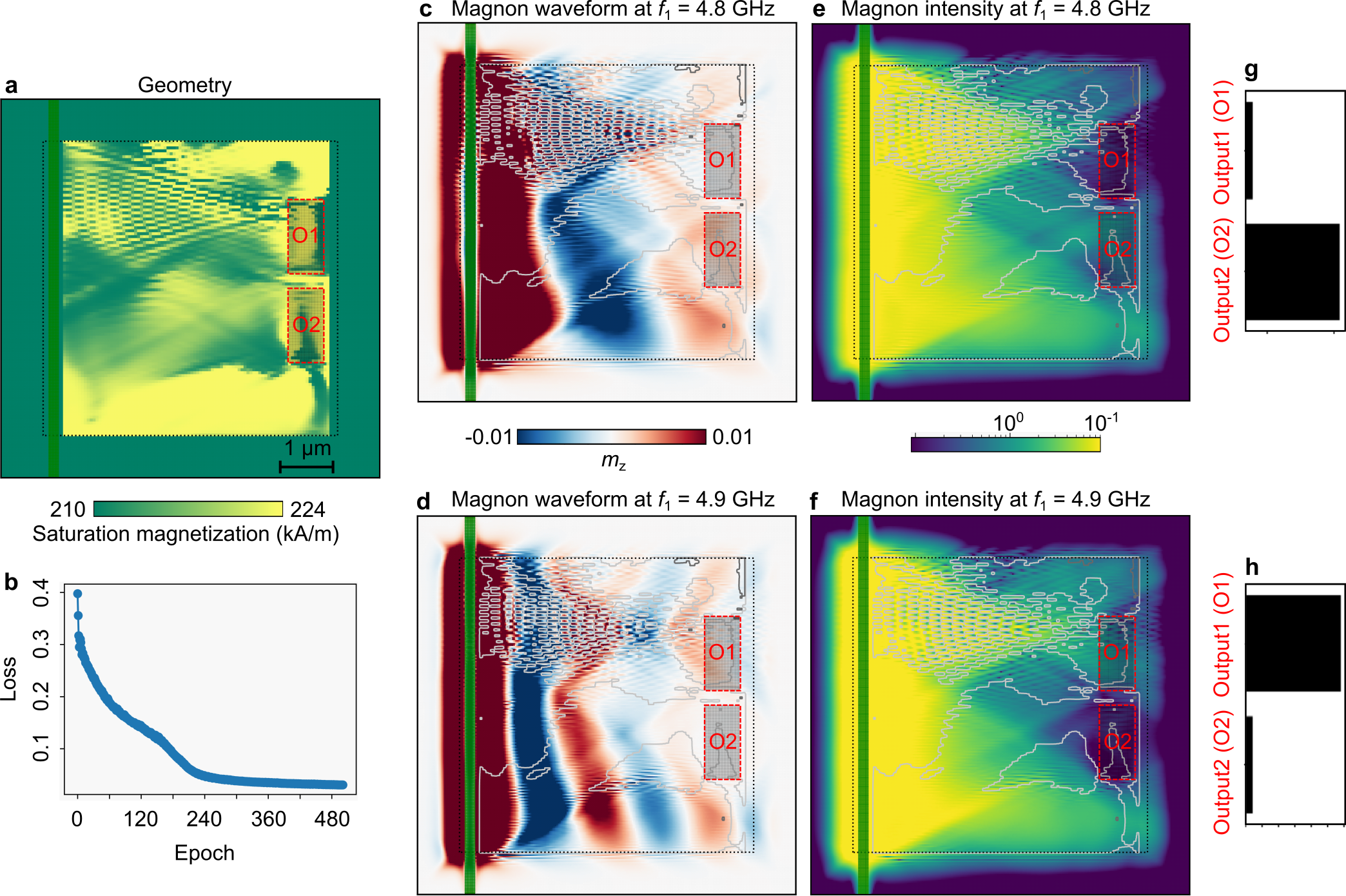}
    \caption{\textbf{Inverse design of a frequency demultiplexer using magnetoelectrically tuned magnon dispersion relations:} \textbf{a} The $M_{eff}$ pattern created by inverse design, used as a scatterer in \textbf{c-f}. The location of the outputs are denoted by semi-transparent rectangles, labeled O1 (Output1) and O2 (Output2). An absorbing boundary layer is applied around the structure. \textbf{b} The loss curve during the gradient descent optimization of the pattern. \textbf{c,d} Snapshot of the spin-wave front emitted from the antenna (green line) and scattered by the inverse designed pattern ($f_1 = 4.8\, \mathrm{GHz}$, $f_2 = 4.9\, \mathrm{GHz}$). \textbf{e,f} Spin-wave intensity in the steady state showing the desired operation, i.e. $f_1$ is focused to O1, and $f_2$ is directed toward O2. \textbf{g,h} The corresponding output levels. }
    \label{Simulation_Waveguides}
\end{figure}

To demonstrate the suitability of multiferroic structures as a basis for advance magnonics devices we inverse designed a  frequency demultiplexer that is able to discriminate between two frequencies for a given ferroelectric domain configuration. For this we used SpinTorch, a micromagnetic simulator with built-in backpropagation optimization capabilities \cite{Papp2021}. In simulation we used the material parameters extracted from the experimental sample, namely $t_{\text{LSMO}}$ = 21 nm, $A_{\text{ex}}$ = 960 nJ/m, $\alpha = 0.006$, $M_{\text{eff}}^{\text{virgin}}$ = 210 kA/m, and a bias field of 80 mT in DE configuration. To model the written state, we used an effective saturation value extracted from the thermal BLS spectrum, $M_{\text{eff}}^{\text{written}}$ = 224 kA/m (i.e. almost $7\%$  change). The inverse-design algorithm was allowed to tune the $M_{\text{eff}}^{\text{virgin}}$ in the design region continuously between the virgin and the written state values, to facilitate the gradient method. We expect that this is also realistic to achieve experimentally as an effective medium since the ferroelectric domain sizes can be written smaller than the typical spin wavelength. The LSMO film was simulated as a single layer, with lateral discretization of 25 nm. Spin waves were generated using a linear source with 150 nm width, approximating the behaviour of a narrow micro-strip line antenna. 

Fig. 4\textbf{a} depicts the final design of the geometry, which is the results of 500 iterations of backpropagation in simulation. The dotted line around the design area delineate an absorbing boundary region around the device, used to eliminate reflections from the sides of the limited simulation area. Fig. 4\textbf{b} shows a monotonic decrease of the loss function during the iterative design process, with a saturation towards the end.
Correspondingly, Fig. 4 \textbf{c,d} display the simulated spin-wave propagation through the designed scatterer for two excitation frequencies. The FE pattern was designed (via its effective equivalent $M_s$ pattern ) such that $f_1$ = 4.8 GHz frequency is focused to output O1, and $f_2$ = 4.9 GHz is directed to O2.  Fig. 4\textbf{e,f} show the time-integrated intensity in the simulated device,  illustrating how wave interference produces the high and low output levels. Finally, Fig. 4 \textbf{g,h} show the corresponding output levels of the simulations obtained by integrating spin-wave intensity in the rectangular areas labeled O1 and O2.
Overall, the presented design demonstrates that multiferroic heterostructures show great potential in nanoscale magnonics device applications. The influence of FE domains on the local dispersion properties is significant enough to allow effective control of spin waves via refraction and diffraction. We note that the design presented above had limitations due to the strongly anisotropic nature of spin waves in the in-plane biased geometry, which causes the formation of strong caustics and limits scattering to very specific angles.

The isofrequency curve for the wavevectors $k_x$ and $k_y$ indicates that the spin waves are scattered with an angle of ~$30^{\circ}$. The shape of the isofrequency contours in k-space depends on the magnetic anisotropy,  and in LSMO grown on NGO (001) exhibits a uniaxial magnetic anisotropy of 45° with respect to the crystal orientation of NGO. Additionally, there is a small out-of-plane anisotropy field which together can explain the "bending" of the spin wave propagation we determined analytically in an earlier work \cite{Merbouche2021}.
\\\\
From a device-design perspective, isotropic spin waves—particularly forward volume waves—would be significantly more versatile for complex architectures and applications. In this study, the use of Brillouin light scattering (BLS), which is insensitive to forward volume waves, constrained us to the in-plane configuration. Future work employing magnetoresistive detection could overcome this limitation and enable direct exploitation of forward volume wave dynamics.
% (ref:  Katrine and Helmut)

\section*{Conclusion}
In summary, our work suggests a new approach to control spin wave propagation by guiding the spin waves along waveguides which are defined solely by controlling the orientation of the FE polarization vector in the BFO layer. The magnetoelectric coupling between the BFO and the LSMO layer induces a difference in the spin wave dispersion relation between these different regions. This shift is reflected by observing a frequency shift of nominally 150 MHz in the thermal magnon spectra, corresponding to a change of $7\,\mathrm\%$ of effective magnetization. 
Correspondingly, the study of RF excited coherent magnons with microfocused BLS reveals indeed the varied spin wavelength (dispersion) along the different FE written magnonic waveguides. 
Our experimental results are well corroborated by dynamic magnetic simulations. Furthermore, using the inverse design method, we showed that our system has indeed potential to be used as new platform for magnonic devices as demonstrated by the simulation of a frequency demultiplexer. 

\section*{Methods}

\textbf{Pulsed laser deposition growth of BFO/LSMO/NGO multilayer}\\
Thin films were grown by RHEED-assisted pulsed laser deposition technique with a KrF excimer laser on single crystal (001)$_\mathrm{o}$-oriented NdGaO$_3$ substrates. Substrates were pretreated prior to deposition by annealing at 1000 $^{\circ}$C for 3 hours in oxygen environment. The deposition process began with the growth of La$_{0.67}$Sr$_{0.33}$MnO$_3$ (LSMO) layer (21 nm) at 750 $^{\circ}$C under an oxygen pressure of 0.4 mbar, followed by the deposition of BiFeO$_3$ (BFO) (34 nm) at 650 $^{\circ}$C under 0.36 mbar. The repetition rates were set at 2 Hz for LSMO and 5 Hz for BFO. The laser fluence was kept at 1 and 1.5 J/cm$^2$ for BFO and LSMO, respectively.

\textbf{Device fabrication}\\
Markers and radio-frequency (RF) antennas were fabricated using electron beam lithography and lift-off techniques. Single-line RF antennas consist of 10 nm-thick Ti and 120 nm-thick Au. All nanofabrication processes were completed before the ferroelectric domain patterning with PFM.

\textbf{Piezoresponse force microscopy imaging}\\
Piezoresponse force microscopy (PFM) images were obtained using an atomic force microscope (Nanoscope V Multimode, Bruker). Two external lock-in amplifiers (SR830, Stanford Research) were used to simultaneously acquire the vertical (VPFM) and lateral (LPFM) responses. For this, an AC voltage at a frequency of 35 kHz was applied between the LSMO (as bottom electrode) and a grounded Pt-coated tip as sketched in Fig. 1\textbf{b}. Cantilevers with stiffness of 40 N$\cdot$m$^{-1}$ were utilized for the measurements. Further, in order to reverse the ferroelectric polarization, DC voltage was applied to bottom electrode by external function generator. 

\textbf{Brillouin light scattering.}\\
Micro-focus Brillouin light scattering ($\mu$BLS)~\cite{Demokritov2007} were conducted at room temperature to detect both the thermal magnons (incoherent) and RF-excited propagating magnons (coherent) in LSMO layer. For both sets of measurements, monochromatic continuous-wave solid-state lasers were used. The laser power of about 1 mW was focused on the BFO/LSMO top surface for thermal magnon detection, and 0.25 mW for RF-excited magnon detections. The recorded BLS signal is proportional to the square of the amplitude of the dynamic magnetization at the position of the laser spot. Back-scattered light was analyzed using the six-pass Fabry-Perot Tandem Interferometer TFP-2 (THE TABLE STABLE LTD.) For the thermal magnon detection in the main text and time-resolved BLS experiment in the Supplementary Information, we utilized the green laser with a wavelength of 532 nm and the laser was focused to a spot of 300 nm diameter using a 100$\times$ objective lens with numerical aperture of NA$~=0.85$. For the propagating magnon detection discussed in the main text, blue laser with wavelength of 473 nm was focused with a 100$\times$ objective lens whose numerical aperture is NA$~=0.85$ and the laser spot size is 300 nm. We measured the spin-wave intensity over two-dimensional regions by scanning the probing laser over the ferroelectric domains and waveguides, with active stabilization ensuring high spatial resolution. Phase information was extracted by interfering the scattered light with the excitation RF current. The processed data yielded $\cos(\phi)$, where $\phi$ is the phase difference between the detected magnon signal at the laser focused spot and the excitation RF source.

\section*{Additional information}
The supplementary information includes S1. Characterization of the epitaxial grown BFO/LSMO and (001)-NGO substrate; S2. Ferromagnetic resonance characterization of the LSMO layer; S3. Cycloid state characterization in BFO layer; S4. k-resolved Brillouin light scattering on LSMO layer; S5. Simulations on magnonic waveguides.

\section*{Author Contributions}
I.B., P.C., and A.A. conceived and designed the experiments. Am.A., S.F., and V. G grew and characterize the growth properties of the devices. I.B., P.C., and K.B. conducted the magnetization characterization and analyzed the data. A.V., S.F., and P.C. designed and conducted the ferroelectric state writing. P.C., S.S., J.-P. A.A, and V. D. conducted the $\mu$BLS experiments and analyzed the data. A.T., and P.C. conducted the $k$-resolved BLS and analyzed the data. A.P. performed micromagnetic simulations for the inverse-design device concept. P.C., A.A., A.P., and I.B. wrote the manuscript with the input from all authors. All authors contributed to the scientific discussion.

\section*{Competing Interests} The authors declare that they have no competing interests.

\section*{Acknowledgments}
We acknowledge the support from Horizon2020 Research Framework Programme of the European Commission under grant no. 899646 (k-NET), ANR TATOO (ANR-21-CE09-0033), and European Horizon 2020 TSAR (No. 964931) projects. A.P. acknowledges support from the Bolyai Janos Fellowship of the Hungarian Academy of Sciences.

\section*{Data availability}
The data that support the findings of this study are available from the corresponding authors upon reasonable request.

\section{Supplementary Information}
\renewcommand{\appendixname}{Supplementary Information}
\renewcommand{\thesection}{\arabic{section}}
\appendix
 \section{Characterization of the epitaxial grown BFO/LSMO and (001)$-$NGO substrate}
\begin{figure}[h]
	\centering
	\includegraphics[width=\columnwidth]{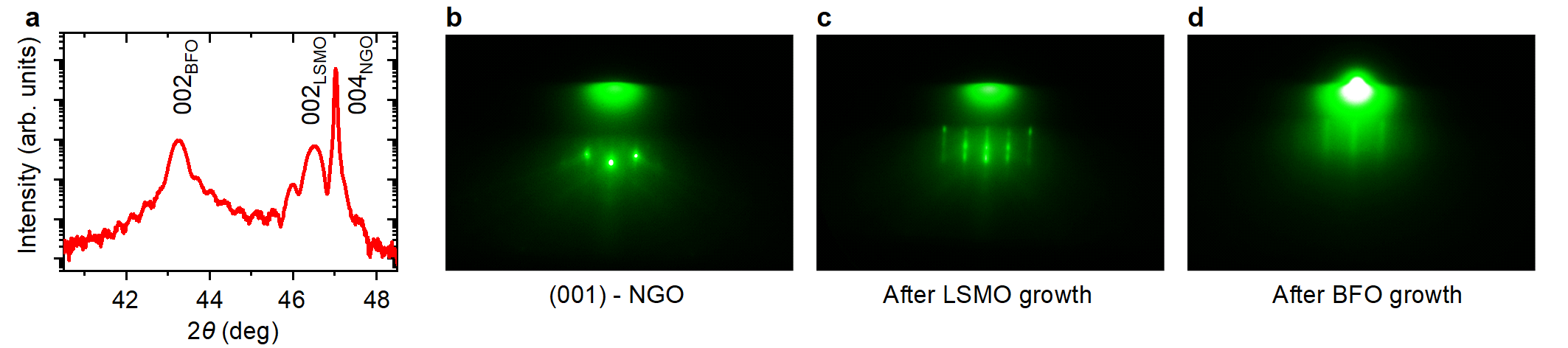}
	\caption{\textbf{XRD and RHEED characterization of BiFeO$_3$ (BFO)/La$_{0.7}$Sr$_{0.3}$MnO$_3$ (LSMO) heterostructure on (001)-NdGaO$_3$ (NGO) substrate.} \textbf{a}, XRD scan of BFO/LSMO heterostructure on (001)$-$NGO substrate. \textbf{b}, \textbf{c}, and \textbf{d}, RHEED intensity during before LSMO layer growth, after LSMO layer growth and after BFO layer growth.}
	\label{XRD_RHEED}
\end{figure}
\clearpage

\section{Cycloid state characterization in BFO layer}

\begin{figure}[h]
	\centering
	\includegraphics[width=0.7\columnwidth]{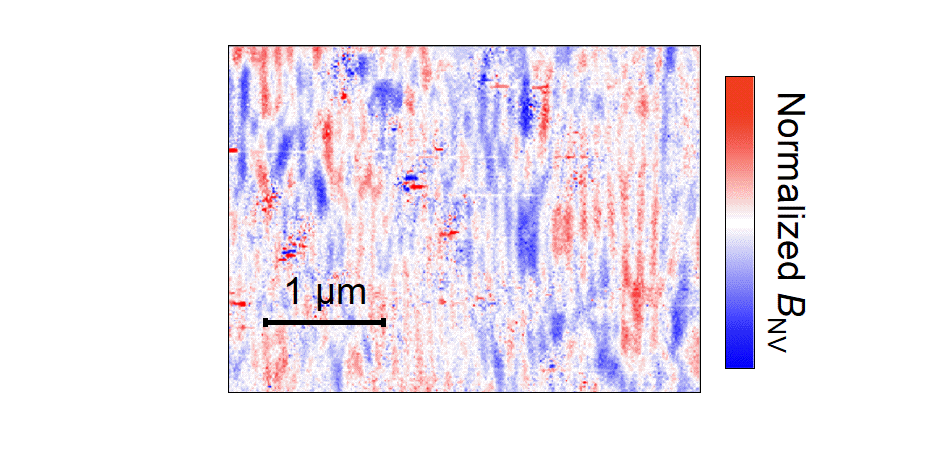}
	\caption{\textbf{Nitrogen-vacancy (NV) centers magnetometry characterizing the cycloid state in BFO layer.} }
	\label{NV}
\end{figure}

\clearpage

\section{Ferromagnetic resonance characterization of the LSMO layer}
\begin{figure}[h]
	\centering
	\includegraphics[width=\columnwidth]{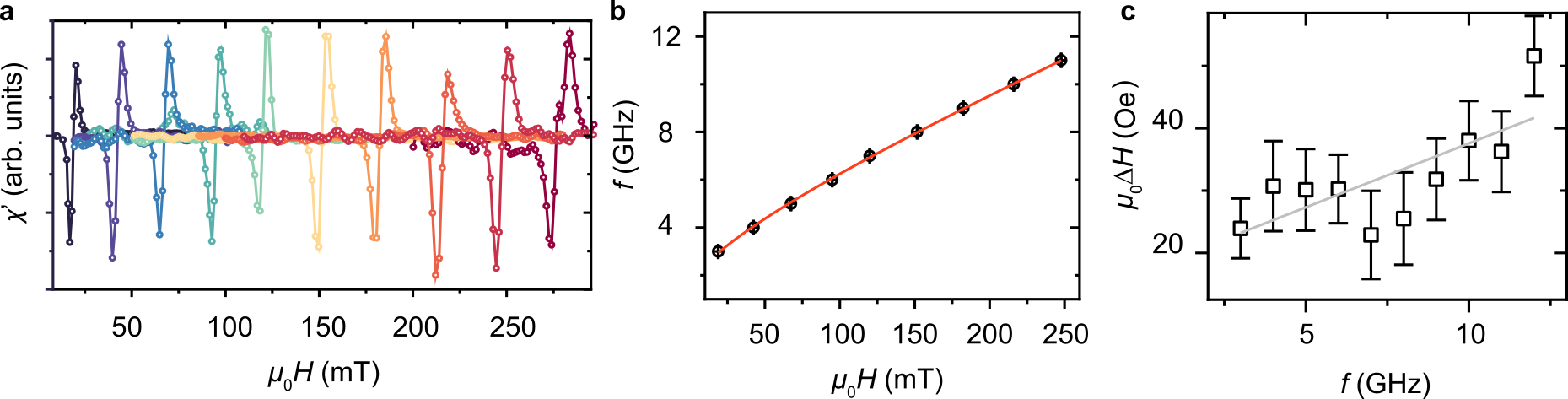}
	\caption{\textbf{Ferromagnetic resonance characterization of the LSMO layer.} \textbf{a} Lineplots of the ferromagnetic resonances detected from 3 GHz (dark blue) to 12 GHz (dark red) from left to right, with 1 GHz as frequency step. \textbf{b} Resonance frequencies extracted via Lorentz shape fitting in \textbf{a}. \textbf{c} Frequency dependent peak-to-peak linewidth reading from \textbf{a} and the linear fitting for damping parameter extraction: $\alpha$ = 5.7 $\times$ 10$^{-3}$.}
	\label{FMR}
\end{figure}

\clearpage

\section{k-resolved Brillouin light scattering on LSMO layer}

\begin{figure}[h]
	\centering
	\includegraphics[width=0.6\columnwidth]{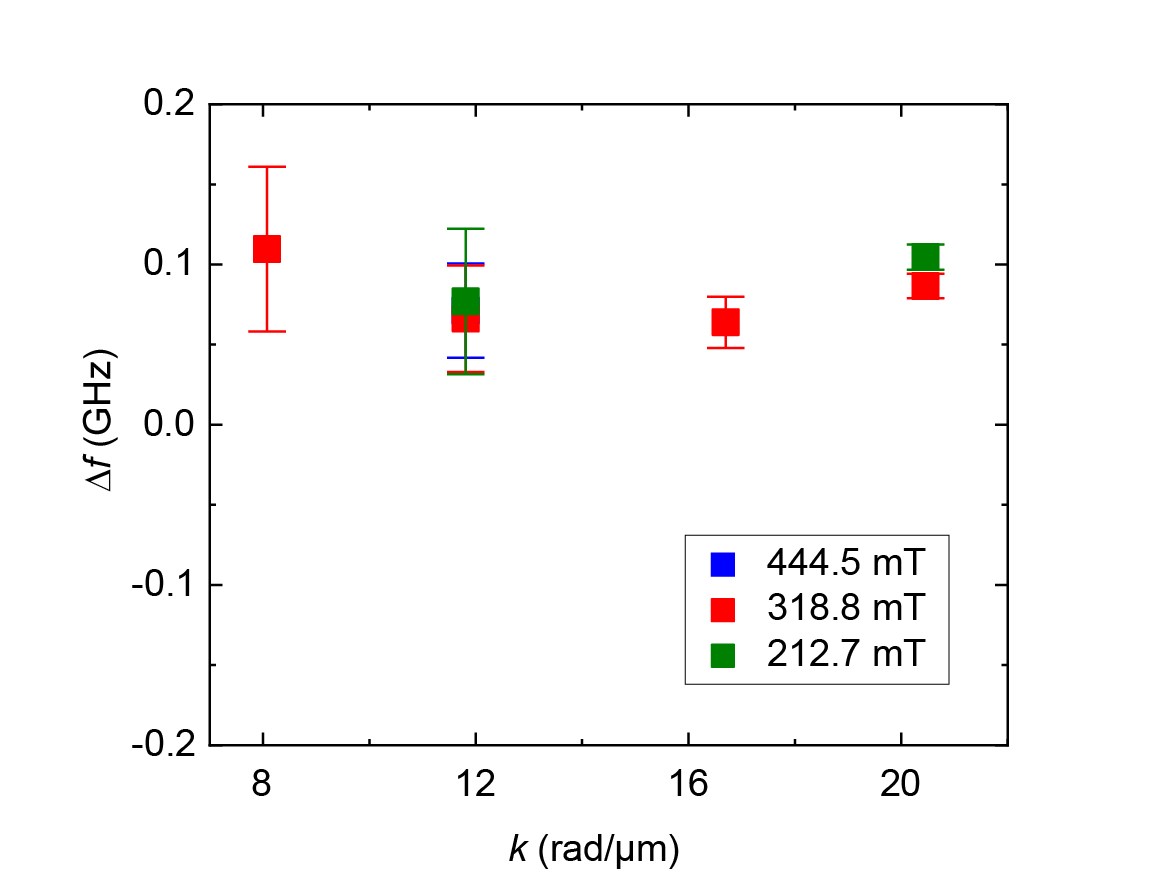}
	\caption{\textbf{Resonance frequency difference characterized by k-resolved Brillouin light scattering (BLS).} }
	\label{k_BLS}
\end{figure}

\clearpage

\section{Simulations on magnonic waveguides}

\begin{figure}[h]
    \centering
    \includegraphics[width=0.6\columnwidth]{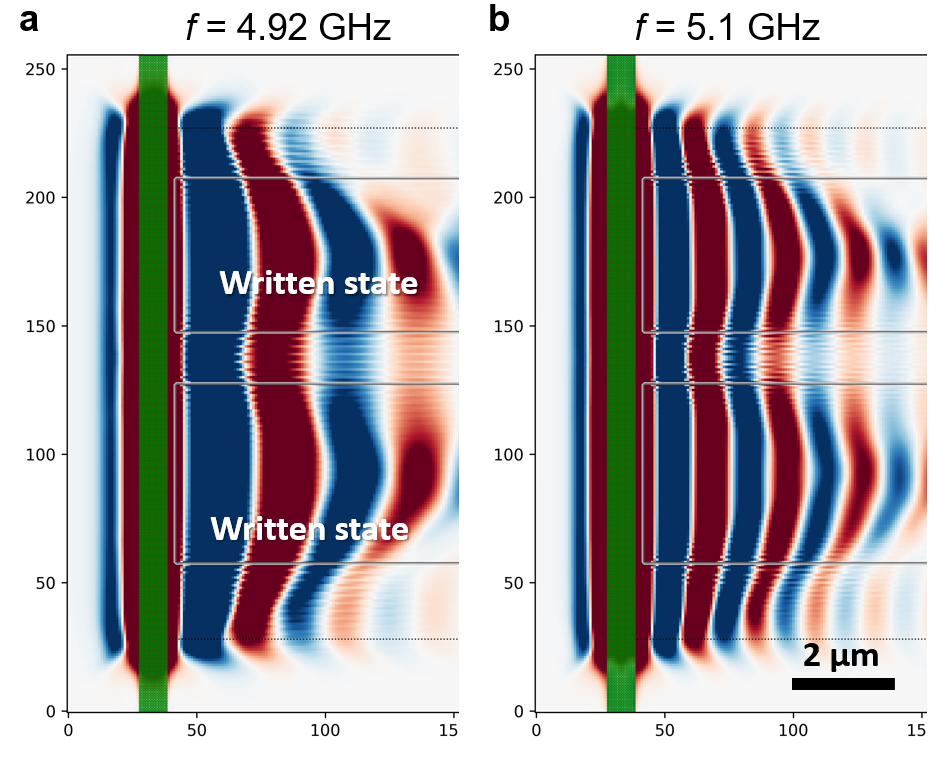}
    \caption{\textbf{Simulated magnon propagation in magnetoelectrically tuned waveguides}}
    \label{Simulation_Waveguides}    
\end{figure}


\begin{thebibliography}{45}	%no more than 50 for Nat. Nanotechnol.

%Reconfigurable magnonic devices
%\bibitem{Krawczyk2014}
%    Krawczyk, M. $\&$ Grundler, D.Review and prospects of magnonic crystals and devices with reprogrammable band structure. \textit{J. Phys.: Condens. Matter} \textbf{26}, 123202 (2014).

\bibitem{Grundler2015}
    Grundler, D. Reconfigurable magnonics heats up. \textit{Nat. Phys.} \textbf{11}, 438–441 (2015).

\bibitem{Chumak2022}
    Chumak, A. V. et al. Advances in magnetics roadmap on spin-wave computing. \textit{IEEE Trans. Magn.} \textbf{58}, 1–72 (2022).

\bibitem{Flebus2025}
    Flebus, B. et al. The 2024 magnonics roadmap. \textit{J. Phys.: Condens. Matter} \textbf{36}, 363501 (2025).

\bibitem{Haldar2016}
    Haldar, A., Kumar, D. $\&$ Adeyeye, A. O. A reconfigurable waveguide for energy-efficient transmission and local manipulation of information in a nanomagnetic device. \textit{Nat. Nanotechnol.} \textbf{11}, 437-443 (2016).

%\bibitem{Wang2017}
%    Wang, Q., Chumak, A. V., Jin, L., Zhang, H., Hillebrands, B. $\&$ Zhong, Z. Voltage-controlled nanoscale reconfigurable magnonic crystal. \textit{Phys. Rev. B} \textbf{95}, 134433 (2017).

\bibitem{Yi2019}
   Yi, D., Yu, P., Chen, Y., Lee, H., He, Q., Chu, Y. and Ramesh, R. , Tailoring Magnetoelectric Coupling in BiFeO3/La0.7Sr0.3MnO3 Heterostructure through the Interface Engineering \textit{Adv. Mat. } \textbf{31}, 1806335 (2019).


\bibitem{Ji2025}
    Ji, H., Li, M., Zhou, G., Guo, Y., Gao, X., Zhou, X. and Liu, L., Recent Progress On Low‐Power Electrical Control of Magnetization in Transition Metal Oxide Heterostructures. \textit{Adv. Funct. Mat.}, 35. (2025)

\bibitem{Wang2018}
    Wang, Q., Pirro, P., Verba, R., Slavin, A., Hillebrands, B. $\&$ Chumak, A. V. Reconfigurable nanoscale spin-wave directional coupler. \textit{Sci. Adv.} \textbf{4}, e1701517 (2018).

\bibitem{Merbouche2021}
    Merbouche, H., Boventer, I., Haspot, V., Fusil, S., Garcia, V., Gouéré, D., Carrétéro, C., Vecchiola, A., Lebrun, R., Bortolotti, P. et al. Voltage-Controlled Reconfigurable Magnonic Crystal at the Sub-micrometer Scale. \textit{ACS Nano} \textbf{15}, 9775–9781 (2021).

\bibitem{Chai2024}
    Chai, Y., Liang, Y., Xiao, C., Wang, Y., Li, B., Jiang, D., Pal, P., Tang, Y., Chen, H., Zhang, Y. et al. Voltage control of multiferroic magnon torque for reconfigurable logic-in-memory. \textit{Nat. Commun.} \textbf{15}, 5975 (2024). 

\bibitem{Zenbaa2025}
    Zenbaa, N., Abert, C., Majcen, F., Kerber, M., Serha, R. O., Knauer, S., Wang, Q., Schrefl, T., Suess, D. $\&$ Chumak, A. V. A universal inverse-design magnonic device. \textit{Nat. Electron.} \textbf{8}, 106–115 (2025).
    
%For the concept of spin waves and magnons for devices
\bibitem{Gurevich1996}
    Gurevich, A. G. $\&$ Melkov, G. A. Magnetization Oscillations and Waves. CRC Press (1996).

\bibitem{Kruglyak2010}
    Kruglyak, V. V., Demokritov, S. O. $\&$ Grundler, D. Magnonics. \textit{J. Phys. D: Appl. Phys.} \textbf{43}, 264001 (2010).

\bibitem{Chumak2015}
    Chumak, A. V., Vasyuchka, V. I., Serga, A. A. $\&$ Hillebrands, B. Magnon spintronics. \textit{Nat. Phys.} \textbf{11}, 453–461 (2015).

\bibitem{Csaba2017}
	Csaba, G., Papp, Á. $\&$ Porod, W. Perspectives of using spin waves for computing and signal processing. {\it Phys. Lett. A} {\bf 381}, 1471–1476 (2017).

\bibitem{Bensmann2025}    
    Bensmann, J. et al. Dispersion-tunable low-loss implanted spin-wave waveguides for large magnonic networks. \textit{Nat. Mater.} \textbf{24}, 1045–1052 (2025).

%Magnon couple with other particles, quasi-particles and states
\bibitem{An2020}%magnon-phonon
    An, K., Litvinenko, A. N., Kohno, R., Fuad, A. A., Naletov, V. V., Vila, L., Ebels, U., de Loubens, G., Hurdequint, H., Beaulieu, N. et al. Coherent long-range transfer of angular momentum between magnon Kittel modes by phonons. \textit{Phys. Rev. B} \textbf{101}, 060407 (2020). 
\bibitem{Guo2018}
    Guo, Er-Jia and Roldan, Manuel A. and Sang, Xiahan and Okamoto, Satoshi and Charlton, Timothy and Ambaye, Haile and Lee, Ho Nyung and Fitzsimmons, Michael R., Influence of chemical composition and crystallographic orientation on the interfacial magnetism in $\mathrm{BiFe}{\mathrm{O}}_{3}$/$\mathrm{L}{\mathrm{a}}_{1\text{\ensuremath{-}}x}\mathrm{S}{\mathrm{r}}_{x}\mathrm{Mn}{\mathrm{O}}_{3}$ superlattices. \textit{Phys. Rev. Mat. } \textbf{2}, 114404 (2018). 
\bibitem{Xu2020}%magnon-phonon
    Xu, M., Yamamoto, K., Puebla, J., Baumgaertl, K., Rana, B., Miura, K., Takahashi, H., Grundler, D., Maekawa, S., Otani, Y. et al. Nonreciprocal surface acoustic wave propagation via magneto-rotation coupling. \textit{Sci. Adv.} \textbf{6}, eabb1724 (2020).

\bibitem{Tabuchi2014}%magnon-photon coupling
    Tabuchi, Y., Ishino, S., Ishikawa, T., Yamazaki, R., Usami, K. $\&$ Nakamura, Y. Hybridizing ferromagnetic magnons and microwave photons in the quantum limit. \textit{Phys. Rev. Lett.} \textbf{113}, 083603 (2014).

%\bibitem{Yao2017}
%    Yao, B., Gui, Y. S., Rao, J. W., Kaur, S., Chen, X. S., Lu, W., Xiao, Y., Guo, H., Marzlin, K.-P. $\&$ Hu, C.-M. Cooperative polariton dynamics in feedback-coupled cavities. \textit{Nat. Commun.} \textbf{8}, 1437 (2017).

\bibitem{Diederich2023}%magnon-photon
    Diederich, G. M., et al. Tunable interaction between excitons and hybridized magnons in a layered semiconductor. \textit{Nat. Nanotechnol.} \textbf{18}, 23–28 (2023).

%\bibitem{Qian2025}%magnon-photon
%    Qian, J., Hong, Q., Wang, Z.-Y., Wu, W.-X., Yang, Y., Hu, C.-M., You, J.-Q., Wang, Y.-P. et al. Unidirectional perfect absorption induced by chiral coupling in spin-momentum locked waveguide magnonics. \textit{Nat. Commun.} \textbf{16}, 8100 (2025).

\bibitem{Tabuchi2015}%magnon-qubit coupling
    Tabuchi, Y., Ishino, S., Noguchi, A., Ishikawa, T., Yamazaki, R., Usami, K. $\&$ Nakamura, Y. Coherent coupling between a ferromagnetic magnon and a superconducting qubit. \textit{Science} \textbf{349}, 405–408 (2015).

\bibitem{Lachance-Quirion2020}
    Lachance-Quirion, D., Wolski, S. P., Tabuchi, Y., Kono, S., Usami, K. $\&$ Nakamura, Y. Entanglement-based single-shot detection of a single magnon with a superconducting qubit. \textit{Science} \textbf{367}, 425–428 (2020).

%Magnonics diode, couplers, multiplexers, conduits, majority gates, spin-wave amplifiers, and transistors
\bibitem{Lan2015}
    Lan, J., Yu, W., Wu, R., $\&$ Xiao, J. Spin-wave diode. \textit{Phys. Rev. X} \textbf{5}, 041049 (2015).
    
%\bibitem{Szulc2020}
%    Szulc, K., Graczyk, P., Mruczkiewicz, M., Gubbiotti, G. $\&$ Krawczyk, M. Spin-wave diode and circulator based on unidirectional coupling. \textit{Phys. Rev. Appl.} \textbf{14}, 034063 (2020).  

\bibitem{Wang2020}
    Wang, Q. et al. A magnonic directional coupler for integrated magnonic half-adders. \textit{Nat. Electron.} \textbf{3}, 765–774 (2020).
    
\bibitem{Heussner2020}
	Heussner, F. et al. Experimental Realization of a Passive Gigahertz Frequency‐Division Demultiplexer for Magnonic Logic Networks. {\it Phys. status solidi – Rapid Res. Lett.} {\bf 14}, 1900695 (2020).

\bibitem{Heinz2020}
    Heinz, B. et al. Propagation of Spin-Wave Packets in Individual Nanosized Yttrium Iron Garnet Magnonic Conduits. \textit{Nano Lett.} \textbf{20}, 4220–4227 (2020).   

\bibitem{Fischer2017}
	Fischer, T. et al. Experimental prototype of a spin-wave majority gate. {\it Appl. Phys. Lett.} {\bf 110}, 152401 (2017).

\bibitem{Talmelli2020}
	Talmelli, G. et al. Reconfigurable submicrometer spin-wave majority gate with electrical transducers. {\it Sci. Adv.} {\bf 6}, eabb4042 (2020).

\bibitem{Chumak2014}
    Chumak, A. V., Serga, A. A. $\&$ Hillebrands, B. Magnon transistor for all-magnon data processing. \textit{Nat. Commun.} \textbf{5}, 4700 (2014).

\bibitem{Wang2024}
	Wang, Y. Z. et al. Voltage-Controlled Magnon Transistor via Tuning Interfacial Exchange Coupling. {\it Phys. Rev. Lett.} {\bf 132}, 076701 (2024).
 		
\bibitem{Cheng2024}
	Cheng, J. et al. A nonvolatile magnon field effect transistor at room temperature. {\it Nat. Commun.} {\bf 15}, 9314 (2024).

%Voltage-control
\bibitem{Rovillain2010}
    Rovillain, P., de Sousa, R., Gallais, Y., Sacuto, A., Méasson, M. A., Colson, D., Forget, A., Bibes, M., Barthélémy, A. $\&$ Cazayous, M. Electric-field control of spin waves at room temperature in multiferroic BiFeO$_3$. \textit{Nat. Mater.} \textbf{9}, 975–979 (2010).

%\bibitem{Liu2011}
%    Liu, T. $\&$ Vignale, G. Electric control of spin currents and spin-wave logic. \textit{Phys. Rev. Lett.} \textbf{106}, 247203 (2011).

\bibitem{Choudhury2020}
    Choudhury, S., Chaurasiya, A. K., Mondal, A. K., Rana, B., Miura, K., Takahashi, H., Otani, Y., $\&$ Barman, A. Voltage controlled on-demand magnonic nanochannels. \textit{Sci. Adv.} \textbf{6}, eaba5457 (2020).

\bibitem{Liu2021}
    Liu, C., Luo, Y., Hong, D., Zhang, S.-S.-L., Saglam, H., Li, Y., Lin, Y., Fisher, B., Pearson, J. E., Jiang, J. et al. Electric-field control of magnon spin currents in an antiferromagnetic insulator. \textit{Sci. Adv.} \textbf{7}, eabg1669 (2021).

\bibitem{Qin2021}
    Qin, H., Dreyer, R., Woltersdorf, G., Taniyama, T. $\&$ van Dijken, S. Electric-Field Control of Propagating Spin Waves by Ferroelectric Domain-Wall Motion in a Multiferroic Heterostructure. \textit{Adv. Mater.} \textbf{33}, 2100646 (2021).

\bibitem{Rana2024}
    Rana, B. $\&$ Otani, Y. Development of Magnonics with Voltage-Controlled Magnetic Anisotropy. In: \textit{Nanomagnets as Dynamical Systems} pp. 71–96. Springer Nanostructure Science and Technology (2024).

%\bibitem{Mishra2024}
%	Mishra, P. K., Sravani, M., Bose, A. $\&$ Bhuktare, S. Voltage-controlled magnetic anisotropy-based spintronic devices for magnetic memory applications: Challenges and perspectives. {\it J. Appl. Phys.} {\bf 135}, 220701 (2024).

%\bibitem{Petrillo2024}
%    Petrillo, A. A. D., Fattouhi, M., Di Pietro, A., Alerany Solé, M., López-Díaz, L., Durin, G., Koopmans, B. $\&$ Lavrijsen, R. Understanding voltage-controlled magnetic anisotropy effect for the manipulation of dipolar-dominated propagating spin waves. Preprint at ${https://arxiv.org/abs/2404.12345}$ (2024).

%Ferroelectric control
\bibitem{Vermeulen2019}
    Vermeulen, B. F., Ciubotaru, F., Popovici, M. I., Swerts, J., Couet, S., Radu, I. P., Lauwaet, K., De Boeck, J., Van Haesendonck, C., Willaert, F. et al. Ferroelectric control of magnetism in ultrathin HfO$_2$/Co/Pt layers. \textit{ACS Appl. Mater. Interfaces} \textbf{11}, 34385–34395 (2019).  
%BFO
\bibitem{Wang2003}%UCB
	Wang, J. et al. Epitaxial BiFeO$_3$ Multiferroic Thin Film Heterostructures. {\it Science} {\bf 299}, 1719–1722 (2003).

\bibitem{Gross2017}
	Gross, I. et al. Real-space imaging of non-collinear antiferromagnetic order with a single-spin magnetometer. {\it Nature} {\bf 549}, 252–256 (2017).

\bibitem{Husain2025}
    Husain, S. et al. Colossal enhancement of spin transmission through magnon confinement in an antiferromagnet. Preprint at ${https://arxiv.org/abs/2503.23724}$ (2025).

%LSMO as a good magnon platform
\bibitem{Liu2019}
    Liu, C., Wu, S., Zhang, J., Chen, J., Ding, J., Ma, J., Zhang, Y., Sun, Y., Tu, S., Wang, H. et al. Current-controlled propagation of spin waves in antiparallel, coupled domains. \textit{Nat. Nanotechnol.} \textbf{14}, 691–697 (2019).  

\bibitem{Zhang2025}
    Zhang, Y. et al. Switchable long-distance propagation of chiral magnonic edge states. {\it Nat. Mater.} {\bf 24}, 69–75 (2025).
    
%BFL/LSMO interface
\bibitem{Wu2010}
	Wu, S. M. et al. Reversible electric control of exchange bias in a multiferroic field-effect device. {\it Nat. Mater.} {\bf 9}, 756–761 (2010).
		
\bibitem{Yu2010}
	Yu, P. et al. Interface ferromagnetism and orbital reconstruction in BiFeO$_3$-La$_0.67$Sr$_0.33$MnO$_3$ heterostructures. {\it Phys. Rev. Lett.} {\bf 105}, 027201 (2010).

%BFO/LSMO of HY's group
\bibitem{Zhang2021}
	Zhang, J. et al. Long decay length of magnon-polarons in BiFeO$_3$/La$_0.67$Sr$_0.33$MnO$_3$ heterostructures. {\it Nat. Commun.} {\bf 12}, 7258 (2021).

%inverse design		
\bibitem{Wang2021}
	Wang, Q., Chumak, A. V. $\&$ Pirro, P. Inverse-design magnonic devices. {\it Nat. Commun.} {\bf 12}, 2636 (2021).
		
\bibitem{Zenbaa2024}
	Zenbaa, N. et al. Realization of inverse-design magnonic logic gates. {\it Sci. Adv.} {\bf 11}, (2025).
		
\bibitem{Papp2021}
	Papp, Á., Porod, W. $\&$ Csaba, G. Nanoscale neural network using non-linear spin-wave interference. {\it Nat. Commun.} {\bf 12}, 6422 (2021).

%\bibitem{neaton2005}
%	Neaton, J. B., Ederer, C., Waghmare, U. V., Spaldin, N. A. \& Rabe, K. M. First-principles study of spontaneous polarization in multiferroic BiFeO$_3$. {\it Phys. Rev. B - Condens. Matter Mater. Phys.} {\bf 71}, 014113 (2005).
		
%\bibitem{finco2022}
%	Finco, A. et al. Imaging Topological Defects in a Noncollinear Antiferromagnet. {\it Phys. Rev. Lett.} {\bf 128}, 187201 (2022).
		
%\bibitem{burns2020}
%	Burns, S. R., Paull, O., Juraszek, J., Nagarajan, V. \& Sando, D. The Experimentalist’s Guide to the Cycloid, or Noncollinear Antiferromagnetism in Epitaxial BiFeO$_3$. {\it Adv. Mater.} {\bf 32}, 2003711 (2020).
		
%\bibitem{dufour2023}
%	Dufour, P. et al. Onset of Multiferroicity in Prototypical Single-Spin Cycloid BiFeO$_3$ Thin Films. {\it Nano Lett.} {\bf 23}, 9073–9079 (2023).
		    
\bibitem{Abdelsamie2024}
	Abdelsamie, A. et al. Interplay between anisotropic strain, ferroelectric, and antiferromagnetic textures in highly compressed BiFeO$_3$ epitaxial thin films. {\it Appl. Phys. Lett.} {\bf 124}, (2024).
		
\bibitem{Kittel}
	Kittel, C. Interpretation of anomalous larmor frequencies in ferromagnetic resonance experiment. {\it Phys. Rev.} {\bf 71}, 270–271 (1947).
		
%\bibitem{Moon2013}
%	Moon, J.-H. et al. Spin-wave propagation in the presence of interfacial Dzyaloshinskii-Moriya interaction. {\it Phys. Rev. B} {\bf 88}, 184404 (2013).
		
%\bibitem{Gladii2016}
%	Gladii, O., Haidar, M., Henry, Y., Kostylev, M. \& Bailleul, M. Frequency nonreciprocity of surface spin wave in permalloy thin films. {\it Phys. Rev. B} {\bf 93}, 054430 (2016).
		
\bibitem{Guo2013}
	Guo, R. et al. Non-volatile memory based on the ferroelectric. {\it Nat. Commun.} {\bf 4}, 1990 (2013).
		
\bibitem{Heron2014}
	Heron, J. T. et al. Deterministic switching of ferromagnetism at room temperature using an electric field. {\it Nature} {\bf 516}, 370–373 (2014).

\bibitem{Rana2019}
    Rana, B. \& Otani, Y. Towards magnonic devices based on voltage-controlled magnetic anisotropy. {\it Commun. Phys.} {\bf 9}, 90 (2019).

\bibitem{Kim2014}
    Kim, Y$-$M. et {\it al.}Direct observation of ferroelectric field effect and vacancy$-$controlled screening at the $BiFeO_3/La_xSr_{1-x}MnO_3$ interface {\it Nat. Mater} {\bf 13}, 1019$-$1025 (2014).

\bibitem{Gradauskeite2024}
Gradauskaite, E. and Yang, C.$-$J. and Efe, I. and Pal, S. and Fiebig, M. and Trassin, M., Magnetoelectric Phase Control at Domain$-$Wall$-$Like Epitaxial Oxide Multilayers { \it Adv. Funct. Mater.}  \bf{35}, 2412831 (2024)

\bibitem{Huang2018}
Huang, B.C. and Yu, P. and Chu, Y. H. and Chang, C.-S. and Ramesh, R. and Dunin-Borkowski, R. E. and Ebert, P. and Chiu, Y.$-$P.,{ Atomically Resolved Electronic States and Correlated Magnetic Order at Termination Engineered Complex Oxide Heterointerfaces},{ \it ACS Nano},\bf{12},1089$-$1095 (2018)

\bibitem{Demokritov2007}
	Demokritov, S. O. \& Demidov, V. E. Micro-brillouin light scattering spectroscopy of magnetic nanostructures. {\it IEEE Trans. Magn.} {\bf 44}, 6–12 (2008). 	
\end{thebibliography}
\end{document}